# The Physics of Protein Self-Assembly

Jennifer J. McManus [1], Patrick Charbonneau [2], Emanuela Zaccarelli [3], Neer Asherie [4*]

[1] Department of Chemistry, Maynooth University, Maynooth, Co. Kildare, Ireland.
[2] Department of Chemistry, Duke University, North Carolina, USA
[3] CNR ISC and Department of Physics, University of Rome, La Sapienza, Italy
[4] Department of Physics and Department of Biology, Yeshiva University, New York, USA.

## Highlights

- Proteins self-assemble into a large variety of structures with different sizes and symmetries
- Several important aspects of protein self-assembly can be understood using coarse-grained models that include the short-range and anisotropic (or patchy) protein-protein interactions
- The ability to predict and design self-assembled structures is limited, though promising approaches exist
- Numerous computational and experimental challenges remain

## Abstract

Understanding protein self-assembly is important for many biological and industrial processes. Proteins can self-assemble into crystals, filaments, gels, and other amorphous aggregates. The final forms include virus capsids and condensed phases associated with diseases, such as amyloid fibrils. Although seemingly different, these assemblies all originate from fundamental protein interactions and are driven by similar thermodynamic and kinetic factors. Here we review recent advances in understanding protein self-assembly through a soft condensed matter perspective with an emphasis on three specific systems: globular proteins, viruses and amyloid fibers. We conclude with a discussion of unanswered questions in the field.

[*] corresponding author

**Graphical Abstract**

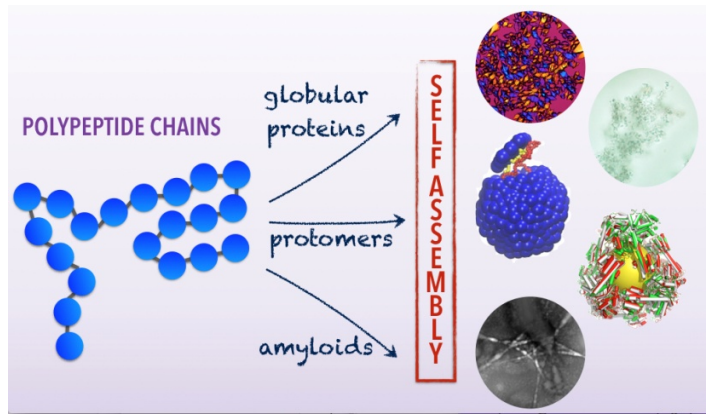

# 1. Introduction

The self-assembly of proteins into small-scale complexes plays a crucial biological role [1]. Under certain conditions, proteins also self-assemble into different structures across a range of length scales from nm to μm (Fig. 1). This process is almost as ubiquitous as complexation, and is equally essential to biology. Some proteins, such as those that make up viral capsids or the outer shell of bacterial microcompartments, self-assemble by design [2, 3]. Others do so when something goes wrong: a conformational change triggers the aggregation of amyloid β-protein (Aβ) into fibrils [4] and a single-point mutation in haemoglobin (Hb) leads to its polymerization [5]. This type of assembly can also result from simple changes to solution conditions (pH, temperature, ionic strength, cosolutes, etc.) [6].

Understanding protein self-assembly is fundamental to many physiological and industrial processes. For example, the fibrillization of Aβ is a feature of Alzheimer's disease [7] and the polymerization of the mutant Hb is the primary pathogenic event of sickle-cell anemia [5]; other protein condensation diseases, for which the pathology is associated with the self-assembly of a condensed protein phase, include cataract formation [8] and Parkinson's disease [9]. In the industrial production of proteins, self-assembly can be harnessed advantageously for protein purification through crystallization or liquid-liquid phase separation [10], or can be problematic if encountered during formulation and storage (often at high protein concentration), when the assembly process is not controlled [11]. Finally, protein self-assembly is essential to structural biology. Most structures are determined through x-ray crystallography, which requires the production of high-quality protein crystals [12].

Here we review recent advances in understanding protein self-assembly. We adopt a soft condensed matter perspective in which simplified models are used to capture the essential elements of protein interactions to determine their assembly. While it is true that atomic-level details are sometimes required for a complete explanation of specific phenomena, the near ubiquitous nature of protein self-assembly suggests the existence of universal elements governing it, which many experimental, computational and theoretical findings support. We focus on three specific systems: globular proteins, viral capsids and amyloid fibrils. These systems formed the core of a 2015 CECAM workshop we organized that brought together researchers from diverse fields (including material science, crystallography, macromolecular chemistry and biophysics) to discuss current challenges in understanding protein self-assembly. This opinion piece builds upon the presentations and discussions at the workshop as well as our own work in the field to stimulate further research—and perhaps breakthroughs—in the physics of protein self-assembly.

## 2. Globular Proteins

Globular proteins have a compact, often spherical shape in which most of the hydrophobic residues are buried in the interior of the protein, while the polar or charged residues are predominantly on the surface in contact with the solvent. Single domain globular proteins range in size from 6 to 300kDa (roughly 60 to 2500 amino acid residues), and multiple domain proteins can be even larger. Globular proteins self-assemble into a variety of states: crystals, dense liquid phases, gels, fibres and amorphous aggregates [13]. A given globular protein can self-assemble into different forms (e.g., crystal vs. aggregate) depending on solution conditions [14]. For some systems, two different states (e.g., two crystal forms) can even coexist [15, 16, 17].

The range of self-assembled states can be summarized in a phase diagram, and for several globular proteins, such as lysozyme [18], γ-crystallins [19] and bovine pancreatic trypsin inhibitor [20], comprehensive phase diagrams have been determined. These phase diagrams share a few common features, the most prominent being that liquid-liquid phase separation is metastable with respect to crystallization (Fig. 2). This metastability can be explained by modelling the globular protein as a simple attractive colloid: a hard spherical core with an isotropic attractive interaction [21]. Numerous theoretical and computational studies have confirmed that for sufficiently short-range attraction (less than about one-quarter the radius of the hard core), liquid-liquid phase separation becomes metastable with respect to crystallization [22], but arrest within the spinodal region of the phase diagram can also occur [23].

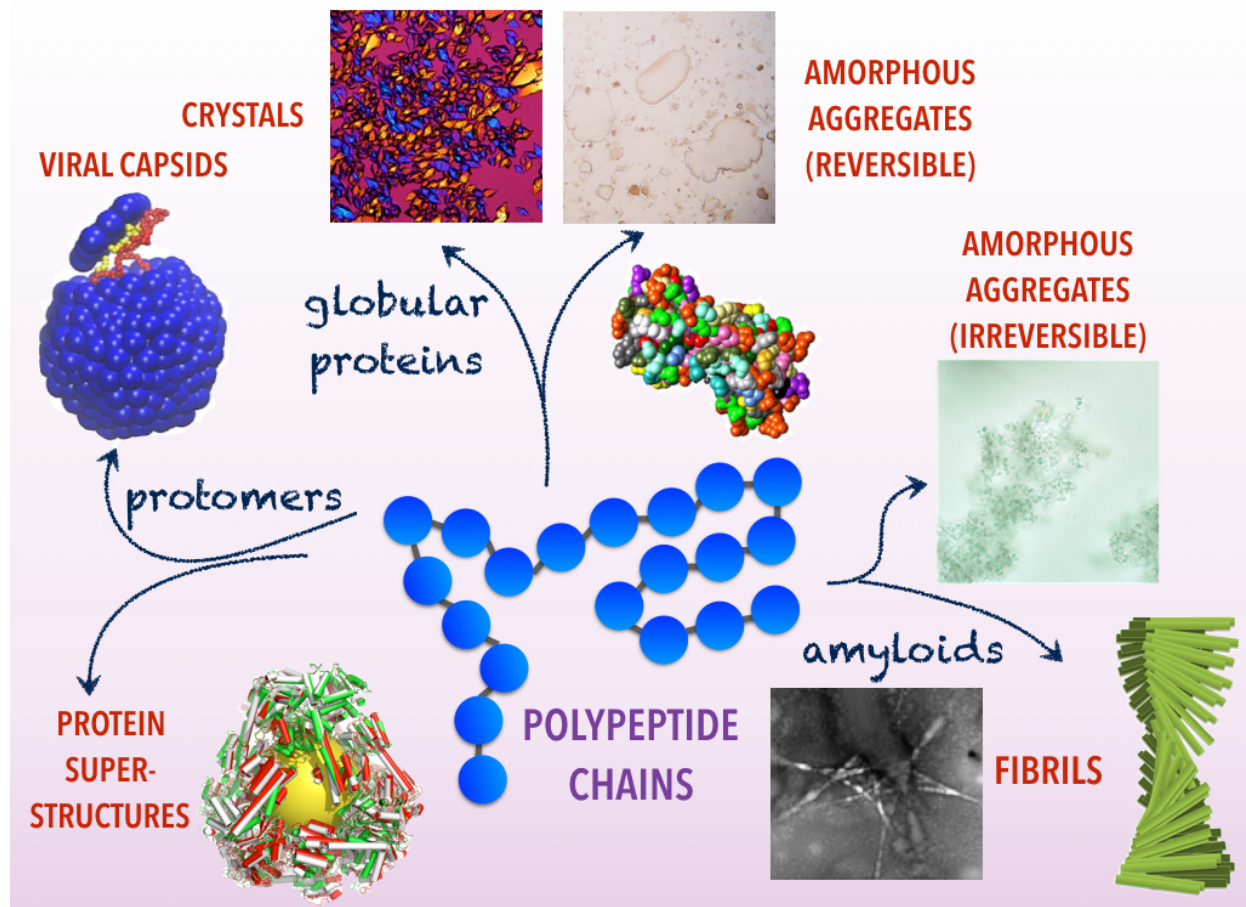

*Figure 1: A number of different assemblies can be formed by either proteins or peptides as illustrated here. Globular proteins can assemble to become either crystals or amorphous aggregates; both types of assembly are reversible (top). Upon a conformational change to a protein or peptide, either amorphous aggregates (which are irreversible) or amyloid fibrils can form (right). Higher order assemblies including viral capsids (in the presence of a nucleic acid) and protein superstructures (including bacterial microcompartments) can also form either naturally or in a directed manner, e.g. by mutagenesis (left).*

The isotropic model does not, however, provide a complete explanation of the self-assembly of globular proteins [24]. Several predictions of isotropic models, such as the shape of the phase boundaries or the crystal density, differ from the experimental observations [25]. While it is possible to modify the isotropic model so that its predictions agree more closely with experiment (e.g., introducing a temperature-dependent energy of interaction or two different ranges of interactions), marked discrepancies remain.

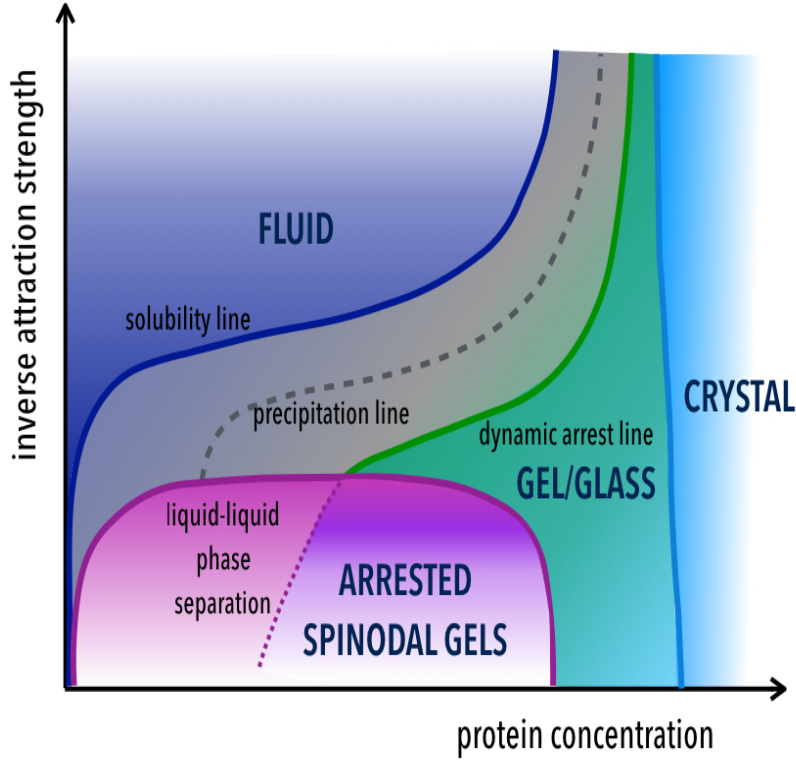

*Figure 2: Illustration of a state diagram for a globular protein indicating the variety of condensed phases that can form. Not all states are present for any given protein and the positions of the boundaries depend on solution conditions. The position of the precipitation line can occur anywhere in the grey region.*

A natural way to improve upon isotropic models is to introduce anisotropy in the protein-protein attraction [26, 27]. This naturally follows from the interaction between surface residues (amino acids) differing strongly from one to another [28]. (To a large extent, this effect dominates over shape anisotropy for simple globular proteins—see conclusions.) A key challenge related to anisotropic models that has yet to be met in a comprehensive fashion is selecting the degree of anisotropy. A common approach consists of placing patches (either randomly or symmetrically) on the surface of a hard sphere and of imposing rules for the interaction between different patches (range, angular width, strength, etc.) [29, 30]. By investigating models with different parameters (including a different number of patches), it is often possible to fit experimental results at hand [31]. While this approach can be effective, it is computationally intensive and must be repeated for each protein and corresponding set of experimental data. It also cannot identify a unique anisotropic model. As a result, it can be challenging to obtain both specific guidance for a given protein and general insights applicable to broad classes of proteins.

Another approach is to use a subset of the experimental results, typically a crystal structure, to determine a specific anisotropic model ("patchiness") and then use this model to analyze other self-assembly data [32]. This approach has not been explored as extensively as the first one, and more work should be carried out in this direction to assess its applicability to many different proteins. A significant limitation is its reliance on the crystal structure. While this information provides information about the patchiness of the protein, it limits the approach to those proteins that have already been crystallized.

Given the importance of protein crystals in structural biology, we believe that one long-term goal should be to develop models for protein self-assembly that help to directly predict crystal and possibly other forms of assembly. At present, protein crystallization is a low-success, brute-force endeavour in which solution conditions are changed or other alterations to the protein interactions are made (e.g., reduction of surface entropy by substituting floppy amino acids with more compact ones) in the hope that crystals are produced [12]. Although various proposals to enhance the likelihood that a protein crystallizes based on our understanding of self-assembly have been made, existing tools remain relatively crude [26, 33].

One notable tool is the Surface Entropy Reduction prediction (SERp) server that, given a protein sequence, suggests sites at which to make mutations that would reduce the surface entropy of the protein and therefore increase the chance the protein will crystallize [34]. It would, however, be useful to have a complementary tool that makes specific predictions regarding crystal self-assembly based on protein interactions. A first step towards this goal would be to devise a method for determining the patchiness of a protein that goes beyond random or symmetric guesses, yet does not require crystallization data a priori. One way to do so would be to identify regions based on the protein sequence and predictions of the protein structure that are likely to be crystal contacts. These regions would then be modeled as attractive or repulsive patches whose properties would be related to the underlying molecular interactions of the regions—a charged patch would be modeled differently from a polar one. A tentative phase diagram could then be calculated for different solution conditions and those that lead to stable crystals forms identified and tested experimentally.

### 3. Viral Capsids

A virus consists of a single copy of a genome (RNA or DNA) inside a closed, protein shell, known as a capsid that protects the nucleic acids. For viruses that encapsulate RNA, capsid self-assembly can occur spontaneously upon mixing protein and nucleic acid, which makes them ideal model systems to study the viral assembly process [35]. The capsid of these viruses is also relatively simple: a single protein molecule thick and often icosahedral (although other polyhedral and helical shapes are also observed) [36]. For example, the 28nm capsid of the cowpea chlorotic mottle virus (CCMV) consists of 180 copies of a single capsid protein (of molecular weight ~20kDa) [37].

As with the assembly of globular proteins, viral capsid assembly is sensitive to solution conditions, in particular pH and ionic strength, and it is not yet possible to fully predict this micelle-like self-assembly [38]. However, unlike many globular proteins, capsids self-assemble by design. The experimental reproducibility and robustness is thus high, making their formation less problematic to analyze. Given the relatively small number of molecules involved, viral capsid assembly is also more amenable to simulations with molecular-scale details than the crystallization of globular proteins, for example. As a result, there is significantly more connection between experiments (which stimulate computational work) and simulations (which lead to experimentally testable predictions) than for other protein self-assembly processes [2].

Minimal models of capsid proteins, like those described earlier for globular proteins, require an interaction anisotropy and specificity to capture qualitative features of their assembly [39]. With this minimal set of features, simulations and experiments show that the assembly of empty capsids proceeds via nucleation and growth, which is analogous to the crystallization of globular proteins [39]. For both processes, the loss of translational entropy during self-assembly is offset by gains from specific hydrophobic, electrostatic, van der Waals and hydrogen-bonding interactions. The non-planar geometry of the capsid does influence the self-assembly process (and may alter the conformation of the proteins involved). Analogous experiments at a larger length scales show that crystallization of colloidal particles is also altered on the surface of a sphere [40].

In the presence of a polynucleotide, the electrostatic interactions between positive charges on the capsid protein and negative charges on the nucleic acids can promote or even dominate the capsid assembly process [2]. The length of the polynucleotide encapsulated is also important, since longer RNA or DNA strands are encapsulated at high entropic cost while shorter strands may not provide sufficient electrostatic stabilization [41]. The optimal nucleic acid length for some viral capsids has been measured [35]. The self-assembly properties of this model are in excellent agreement with experimental results for a variety of viruses.

An ongoing challenge related to viral self-assembly is the characterization of the interaction anisotropy that results in both protein-protein and protein-genome interactions. A further challenge is to identify and characterize intermediate structures along the assembly pathways. It is interesting to note that a similar challenge is faced in the study of protein crystallization, where two main ordering pathways have been proposed: direct nucleation from a solution of monomers, and the formation of long-lived metastable protein clusters from which a crystal emerges [42].

## 4. Amyloid fibrils

Amyloid fibrils are insoluble protein aggregates with a cross-β structure in which β-strands form almost continuous hydrogen-bonded β-sheets that run along the fibril. The fibrils formed from different proteins are qualitatively similar: unbranched filamentous structures that are a few nanometers in diameter, but can grow to be several microns in length. This common structure is likely driven by the universal tendency of polypeptide chains to form hydrogen bonds between atoms along the backbone [43]. It has been suggested that the fibril can be thought of a one-dimensional pseudo-crystal and in that sense, the fibril is the most organized structure that a flexible polypeptide chain can form. It has also been speculated that under certain conditions, an amyloid fibril made of short polypeptides (< 150 residues) would be thermodynamically more stable than even the functional native state [43]. This proposal could help explain why most amyloid-related diseases are caused by short peptides or proteins. It also implies that understanding the kinetics of amyloid formation is essential for disease prevention.

The most-studied protein that forms amyloid fibril is amyloid β (Aβ), a peptide of 40-42 amino acids associated with the pathogenesis of Alzheimer's disease [44]. A combination of experiments and simulation indicate that, just as for protein crystallization and viral capsid assembly, there are two main pathways to form the condensed phase: the fibrils may directly nucleate from a solution of Aβ monomers or first self-assemble into an oligomer from which the fibril forms in a second nucleation event [7]. In both cases, there is a conformation change that occurs as the fibril forms. In addition, existing fibrils can seed the formation of new fibrils that then subsequently break off from the original fibril [7].

When fully atomistic models are used, it is possible to obtain reasonable agreement between simulations and experiments of Aβ fibrilization [45]. Coarse grained models based on patchy particles have also been described [46, 47, 48, 49]. However, the factors that control the specific pathway that is observed under a given set of conditions remain poorly understood. The fibrilization is strongly dependent on ionic strength and the identity of the peptide, and it is currently not possible to predict the pathway that will be taken by a specific peptide under a given set of conditions [50].

Despite considerable effort, a number of issues remain unresolved. It is increasingly accepted that for Aβ, oligomeric peptide assemblies are more neurotoxic than fibril plaques, yet the links between the kinetics and mechanisms for fibril growth in-vitro and those in-vivo are not always clear [44]. For Aβ and other amyloid forming peptides and proteins, cell membranes may be important in the nucleation process [51, 52]. Since much of the work on amyloid-forming peptides and proteins is driven by a search for therapies to prevent, slow down or even stop the course of amyloid-related diseases, there is a pressing need to develop a complete picture of the process of self-assembly for amyloid forming proteins and peptides.

## 5. Conclusions and future directions

Protein self-assembly plays an important role in numerous biophysical process. Here we selected three systems—globular proteins, viral capsids and amyloid fibrils—to illustrate recent progress and current challenges in understanding protein self-assembly from a soft matter viewpoint. We conclude by highlighting several topics that in our opinion should be further investigated.

*Solvent effects:* A comprehensive understanding of the role of the solvent in protein self-assembly has yet to be achieved. In aqueous solution, the role of pH and ionic strength has been considered for many proteins and explained using both isotropic and anisotropic colloidal models [53, 54, 55]. The addition of other solvents (and small molecules) such as glycerol, carbohydrates, amino acids and nucleic acids can have a profound influence on protein self assembly (either by direct interaction with the protein surface, or by modifying the solvent characteristics) [12, 43, 56]. These effects are particularly important when formulating proteins for liquid storage or lyophilisation and further understanding of the mechanisms for self-assembly would improve the success of these processes [11]. Crystallization can be improved by the addition of glycerol, which can suppress nucleation resulting in the formation of fewer, larger crystals [57].

Simulations are a natural way to analyze solvent effects as it is possible to treat the solvent explicitly. Given the computational cost of working with explicit solvent, a multiscale approach is usually used: two proteins are simulated in a solvent to deduce the parameters of an effective pair interaction energy that is then used to study self-assembly in a solvent-free system [47]. Another approach uses a coarse-grained optimized potential for efficient protein structure prediction (OPEP) combined with hydrodynamics [58] and may be a useful computational tool in the future. There are cases, however, for which it is not possible to separate out the solvent without sacrificing essential details, such as for membrane proteins, which can only exist in their native state when embedded in a lipid membrane or solubilized in a detergent. Not surprisingly, our understanding of the self-assembly of membrane proteins is still in its infancy.

*Small molecules:* The conjugation of small molecules to proteins is routinely performed to conduct analytical testing, for example by fluorescent tagging, to improve the biological compatibility, for example by PEGylation or to develop a new therapeutic product, for example in a protein-drug conjugate [59, 60, 61]. In each case, a small molecule is covalently attached to a protein to modify its behavior. This in turn can alter the self-assembly [59]. In most cases, conjugation occurs at either a primary amine or at a free cysteine, but specific chemistry to modify proteins at other amino acids has been developed [62]. While these strategies are often used, there are far fewer experimental, simulation or theoretical descriptions of how these modifications alter self-assembly than for unmodified proteins.

*Non-compact proteins:* When modeling proteins for coarse-grained simulations of self-assembly, it is usually assumed that shape anisotropy is negligible, with exceptions only when the anisotropy is too pronounced (e.g., spherocylinders to model Aβ [47]). Although this may be a reasonable assumption for most single-domain proteins, for multi-domain proteins it must be used with care, and there are families of proteins, for which the isotropic-shape approximation is a gross oversimplification.

*Kinetics:* Much of the work on protein self-assembly has focused on the thermodynamic behavior, but kinetics often play a crucial role in determining the outcome of assembly [63, 64,

65]. The most notorious example is that of protein crystallization. Although the crystal may be the most stable state under the conditions studied (as can be verified by seeding the solution with a crystal and watching it grow), a crystal may not form spontaneously, even after many months, if the nucleation rate is slow [66, 67]. In order to achieve the goal of a comprehensive understanding of protein self-assembly (and in particular, the ability to predict what will happen for a given set of conditions), additional computational work should be carried out to examine the kinetics of self-assembly for particles with anisotropic interactions and connect the results with experimental data on nucleation and growth of self-assembled phases.

*Purposeful vs. incidental self-assembly:* As we mentioned in the introduction, some proteins self-assemble by design while others do so only when things go wrong or the solution conditions are perturbed. In other words, the structure of some proteins is such that self-assembly occurs for a specific purpose (such as to encapsulate other molecules), while for other proteins self-assembly is not central to their function—which they carry out in the unassembled state—and only occurs because of incidental physical considerations. It would be interest to compare the properties of proteins involved in purposeful and incidental processes to see whether any new insights may be gleaned regarding the specific and universal features of protein self-assembly (Fig. 3).

We hope that this brief overview of the physics of protein self-assembly will stimulate others to tackle some of the outstanding issues in the field.

**Future Directions**

- A more comprehensive view of the kinetics and intermediate pathways of protein self-assembly.
- Minimal models for proteins self-assembly that do not require crystal structures as experimental inputs.
- A better understanding of how molecular anisotropy directs protein assembly.
- Greater use of atomistic models with explicit solvent to describe protein-protein interactions.

| MODEL DESCRIPTION | MODEL ILLUSTRATION | TYPES OF TACKLED PROTEIN ASSEMBLIES | | | |
|---|---|---|---|---|---|
| | | GLOBULAR | AMYLOIDS | VIRUSES | MEMBRANE |
| ISOTROPIC ATTRACTION | 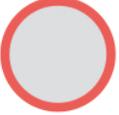 | YES | YES | YES | YES |
| PATCHY SPHERE | 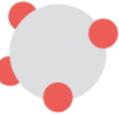 | YES | YES | YES | NO |
| ONE-TO-ONE PATCHY | 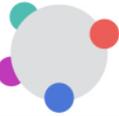 | YES | YES | NO | NO |
| PATCHY ROD/ELLIPSOID | 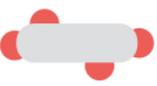 | YES | YES | NO | NO |
| ATOMISTIC WITH IMPLICIT SOLVENT | 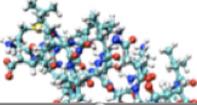 | YES | YES | YES | YES |
| ATOMISTIC WITH EXPLICIT SOLVENT | 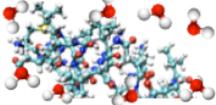 | NO | NO | NO | YES |

*Figure 3: The approaches taken to model protein self-assembly, from simple isotropic models to all atom simulations with explicit solvent, have been used to different extents to support or explain experimental data. With increasing system size, the feasibility of conducting all-atom simulations with explicit solvent reduces significantly.*

**Acknowledgements**

We thank the participants in the CECAM workshop "The Physics of Protein Self-Assembly" (Lausanne, Switzerland, June 2015) for their insightful presentations on the many subjects related to the topic of this article. We thank Michael Hagan and Todd Yeates for permission to use illustrations they created. We gratefully acknowledge financial support from CECAM and the National Science Foundation (DMR 1505287). JM gratefully acknowledges support from Science Foundation Ireland Stokes Lectureship and grant (11/RFP.1/PHY/3165). PC gratefully acknowledges support from the National Science Foundation (DMR-1055586). EZ gratefully acknowledges support from MIUR Futuro in Ricerca ANISOFT (RBFR125H0M). NA gratefully acknowledges financial support from the National Science Foundation (DMR 1206416).